\documentclass{mem}
\usepackage{natbib}\usepackage{txfonts}\usepackage{balance}
\usepackage{graphicx}
\usepackage[a4paper]{hyperref}
\idline{75}{282}
\begin{document}
\def\teff{$T\rm_{eff }$}
\def\kms{$\mathrm {km s}^{-1}$}

\title{Magnetic fields and shock waves in cluster outskirts 
}

   \subtitle{}

\author{
M.~Br\"uggen\inst{1} 
\and R.~J. van Weeren\inst{2}
\and H.~J.~A. R\"ottgering\inst{2}
}
\offprints{M.Br\"uggen}

\institute{
Jacobs University Bremen,
Campus Ring 1
28759 Bremen
Germany
\email{m.brueggen@jacobs-university.de}
\and Leiden Observatory, Leiden University, P.O. Box 9513, NL-2300 RA Leiden, The Netherlands
}

\authorrunning{Br\"uggen}

\titlerunning{Cluster Outskirts}

\abstract{Both, X-ray and radio observations, are beginning to probe regions in galaxy clusters close to the virial radius. These are particularly interesting regions as they are (i) crucial for the thermal evolution of the intra-cluster medium and (ii) they host plasmas under conditions that are poorly understood. Recently discovered giant radio relics in the galaxy cluster CIZA~J2242.8+5301 provide unprecedented evidence for particle acceleration in cluster outskirts. We will discuss some of the intriguing puzzles that surround these relics. The polarisation of the radio emission yields information about the topology of magnetic fields in these shock waves. Here we report on MHD simulations that highlight some problems about the origin of these magnetic fields.

}
\maketitle{}

\section{Introduction}
Radio relics are diffuse radio sources that are usually located in cluster outskirts and that are characterized by steep
radio spectra\footnote{The spectrum is defined as $S(\nu) \propto
  \nu^{- \alpha}$.} with $\alpha \geq 1.2$. The origin of the relativistic electrons - that produce the synchrotron radiation in the radio band - in these very extended sources is still not entirely clear. In the past two years, there has been quite a bit of new data on radio relics with the discovery of the relic in the galaxy cluster CIZA~J2242.8+5301 being the most prominent. CIZA~J2242.8+5301 is a disturbed galaxy cluster located at $z=0.1921$ \citep{2007ApJ...662..224K}. The cluster hosts a large double radio relic system, as well as additional large-scale diffuse radio emission \citep{2010Sci...330..347V}.This recent data has given support to models that explain relics through diffusive particle acceleration at shock fronts, e.g.  \citep{1998A&A...332..395E, 2001ApJ...562..233M}.  The spectral index ($\alpha$) across the bright northern relic steepens systemically in the direction of the cluster center, across the full length of the narrow relic. This is expected for outwards moving shock waves, with synchrotron and inverse Compton losses behind the shock front. However, there are still quite a few puzzles that surround radio relics, of which I will list the most intriguing:

\begin{itemize}
\item Which processes accelerate electrons so efficiently at relatively low Mach number ($M\sim 2-4$) shocks?
\item What produces the magnetic fields inside relics? Both the inferred field strengths as well as the observed polarisation of the radio emission need to be explained.
\item Why do some relics have very sharp edges while others appear very fuzzy?
\item Under which conditions do relics form? When do we see single and when double relics?
\item Some relics appear to be connected to cluster-wide radio halo emission. Is there a physical connection between the two?
\end{itemize}

Many of these questions touch physical processes that are poorly understood, such as diffusive shock acceleration and magnetogenesis in collisionless shocks, and relate to regions of the cosmos, namely clusters outskirts, of which we know very little. Ongoing work with radio interferometers, in particular with the new LOFAR radio telescope is likely to bring great advances in these areas.

In these proceedings, we are going to present some recent theoretical work that addresses the orientation of magnetic fields in cluster outskirts. 

\section{Simulations of mergers}

Idealised simulations of clusters have been presented by \cite{roettiger93, schindler93,pearce94, roettiger97}.
\citet{roettiger93} simulated the head-on collision of galaxy clusters with unequal mass, each of them modelled as King-spheres. They used the ZEUS code with a non-uniform grid and the Hernquist (1987) treecode for the dark matter. The simulations showed a bar of X-ray emission perpendicular to the collision axis that coincided with a shock of $M \sim 4$.
\citet{schindler93} used a uniform grid-PPM code + an N-body code to study the head-on collision of galaxy clusters. They observed several shocks that lead to a substantial enhancement in X-ray luminosity (up to a factor of 2). Their simulation did not include cooling. Later on, \citet{burns94} simulated the collision between the Coma cluster and the group NGC 4839, using the ZEUS code with radiative cooling, and obtained similar results.

None of these works included the baryonic contribution to the gravitational potential.

\citet{pearce94} simulated the head-on collision between two equal-mass clusters using smoothed-particle hydrodynamics (SPH) that includes, both, baryons and dark matter.  They found that the baryons end up in extended constant density cores after the collision, whereas the dark matter does not.
\citet{roettiger97} used a PPM/particle-mesh code to simulate the radio relic in A754, varying the impact parameter and mass ratios of the collision.
\citet{ricker98} presented simulations of off-centre collisions between clusters using a PPM code with an isolated multigrid solver for the gravitational potential. They studied how the virialisation time, the structure of the merger and the X-ray luminosity enhancement depends on the impact parameter of the collision. They find that the X-ray luminosity increases by a large factor, which strongly depends on the impact parameter, and lasts for about half a sound crossing time after the cores interact. Virialisation takes about 5 sound crossing times.
\citet{ricker01} presented a parameter study of off-centre collisions between galaxy clusters of two different mass ratios. Using the Eulerian hydrodynamics/N-body COSMOS code, they collide hydrostatic clusters with a $\beta$ model gas profile. They note that the jumps in X-ray luminosity and temperature have important consequences for cosmological parameters derived from the observations of hot clusters at high redshift. \citet{ricker01} also find that mergers lead to large-scale turbulent motions with eddy sizes of several hundred kpc and this turbulence is driven by dark matter-driven oscillations in the gravitational potential.

\subsection{Method}

The simulations were performed with the multidimensional adaptive mesh
refinement hydrodynamics code FLASH code (\citealt{dubey09})
version 3.2.

A directionally unsplit staggered mesh algorithm (USM) that solves
ideal and non-ideal MHD governing equations in multiple dimensions is
newly implemented in this latest version of the code.  The overall
procedure of the USM scheme can be broken up is described in the FLASH
manual and in Lee (2006) and Lee and Deane (2007).

We model the ICM as ideal gas in hydrostatic equilibrium. The density follows a $\beta$ profile

\begin{equation}
\rho (r) = \rho_0 [1+(r/r_0)^2]^{-3\beta/2} ,
\end{equation}
where $\beta=0.3$, $\rho_0=10^{-25}$ g/cm$^3$ and $r_0=70$ kpc are parameters.

The temperature profile is taken to be
\begin{equation}
T(r) = (T_0+ar)\,[1-(1+\exp(-(r-r_b)/b))^{-1}] ,
\end{equation}
where the parameters have been chosen to be $T_0=3\times 10^7$ K, $a=-5.6$ K/pc, $b=10$ kpc and $r_b=27$ kpc.

The gravitational potential follows from the assumption of hydrostatic equilibrium.

\begin{equation}
\frac{d\phi}{dr} = \frac{k_B}{\mu m_H}\left ( T \frac{1}{\rho}\frac{d\rho}{dr} +\frac{dT}{dr}\right )
\end{equation}

We move the centre of the gravitational potential of cluster 1 on a trajectory around the fixed potential of cluster 2 and thus ignore all interactions between dark matter.

Furthermore, the hydrogen number density was assumed to be related to the electron number density as $n_{H}=n_{\rm e}/1.2$.
We did not include the effects of radiative cooling or star formation, as these are not very relevant for the spatial and temporal time scales considered here.

We set up an initially tangled magnetic field that follows the thermal component according to:
\begin{equation}
\langle {\bf B}\rangle(r)= \langle {\bf B}_0\rangle \left( \frac{n_e(r)}{n_0} \right)^{\eta}
\label{eq:BProfile}
\end{equation}
where $\langle {\bf B}_0\rangle$ is the mean magnetic field strength at the
cluster center and $n_0$ is the number density at the cluster centre.

Simulations start considering a power spectrum for the vector potential ${\bf A}$ in the
Fourier domain:
\begin{equation}
|A_k|^2 \propto k^{-\zeta}
\end{equation}
and extract random values of its amplitude $A$ and phase $\phi$. $A$
is randomly extracted from a Rayleigh distribution (in order to
obtain a Gaussian distribution for the real magnetic field components,
as frequently observed), while $\phi$ varies randomly from 0 to 2$\pi$.
All three components of magnetic field were treated independently which ensures that the final distribution of $B(r)$ has random phase.

The magnetic field components in the Fourier space are then obtained
by:
\begin{equation}
{\tilde B}(k) =ik\times {\tilde A}(k).
\end{equation}

The field components $B_i$ in the real space are then derived using a 3D
Fast Fourier Transform. The resulting magnetic field has the following properties:\\ 
1)$$\nabla \cdot {\bf B}=0,$$\\ 
2) the magnetic field energy density associated with each component
$B_k$ is:$$|B_k|^2 = C^2_n k^{-n},$$ $n= \zeta -2$, where $C^2_n$ is
the power spectrum normalization,\\ 
3)$B_i$ has a Gaussian
distribution, with $\langle B_i\rangle=0$, $\sigma_{B_i}=\langle
B^2_i\rangle$,\\ 
4) $B$ has a Maxwellian distribution.\\ 

For $n=11/3$, the above prescription would give a Kolmogorov distribution of magnetic fields.

One can define $\Lambda=\frac{2\pi}{k}$ as the
physical scale of the magnetic field fluctuations in the real space.
Thus in order to determine the magnetic field power spectrum in the
cluster, we have to determine three parameters: $\Lambda_{min}$,
$\Lambda_{max}$ and $n$.

There are several indications that the magnetic field intensity
decreases going from the center to the periphery of a cluster. This is
expected by magneto-hydrodynamical simulations (see e.g. Dolag et
al. 2008) and by spatial correlations found in some clusters between
thermal and non-thermal energy densities (Govoni et al. 2001). \\ 

 \section{Results}

Here we only have space to show a small part of our results. For further results we refer to an upcoming publication (Br\"uggen, Lee, van Weeren, R\"ottgering in prep.). In Fig.~1 we show our initial conditions and in Fig.~2 we show slices of the gas density and magnetic field vectors in a head-on collision of two galaxy clusters at various times. These are quite representative of many of our simulations. Our main findings can be summarised as follows:

\begin{figure}
\centering{\includegraphics[width={0.49\columnwidth}]{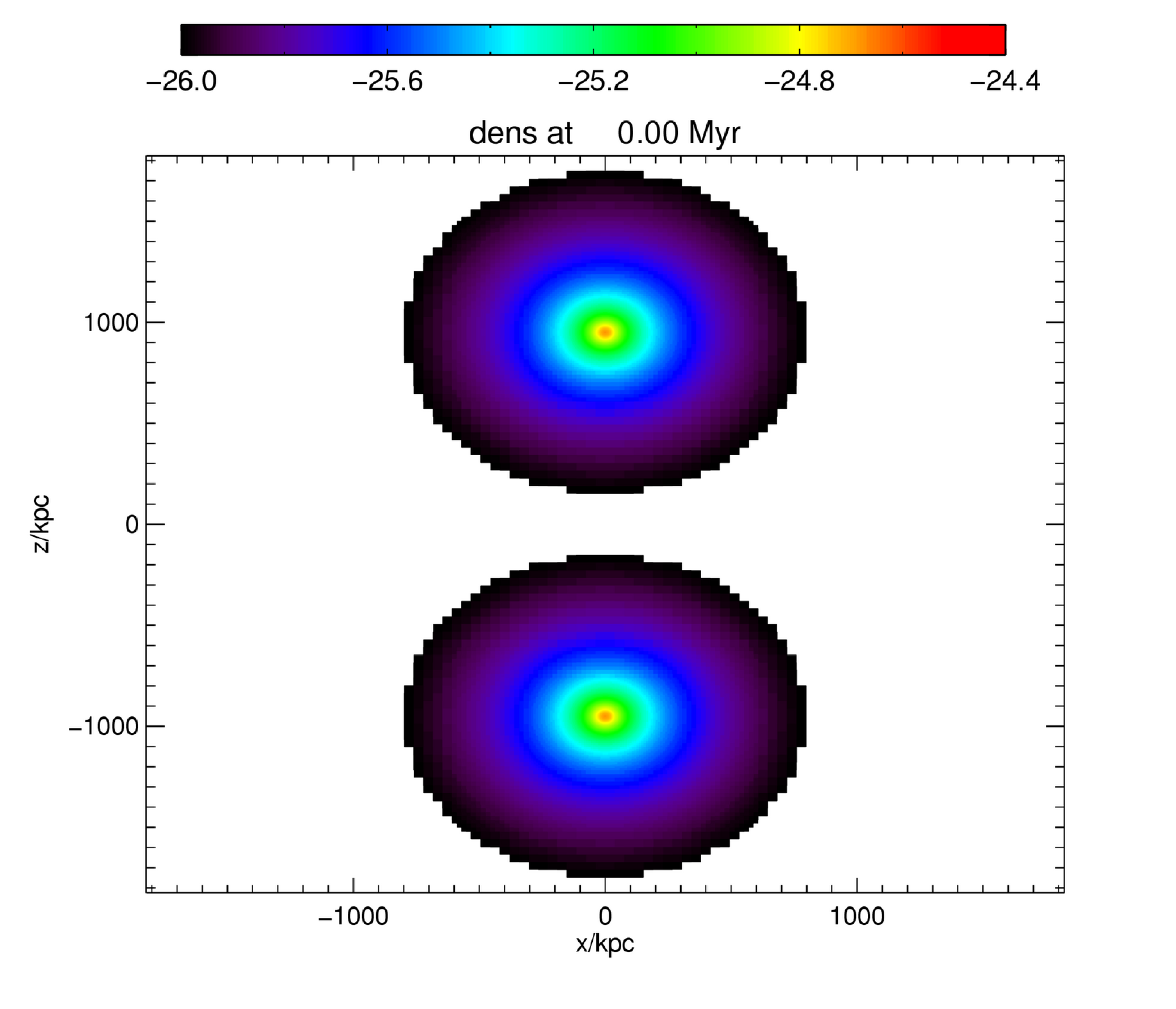}}
\centering{\includegraphics[width={0.49\columnwidth}]{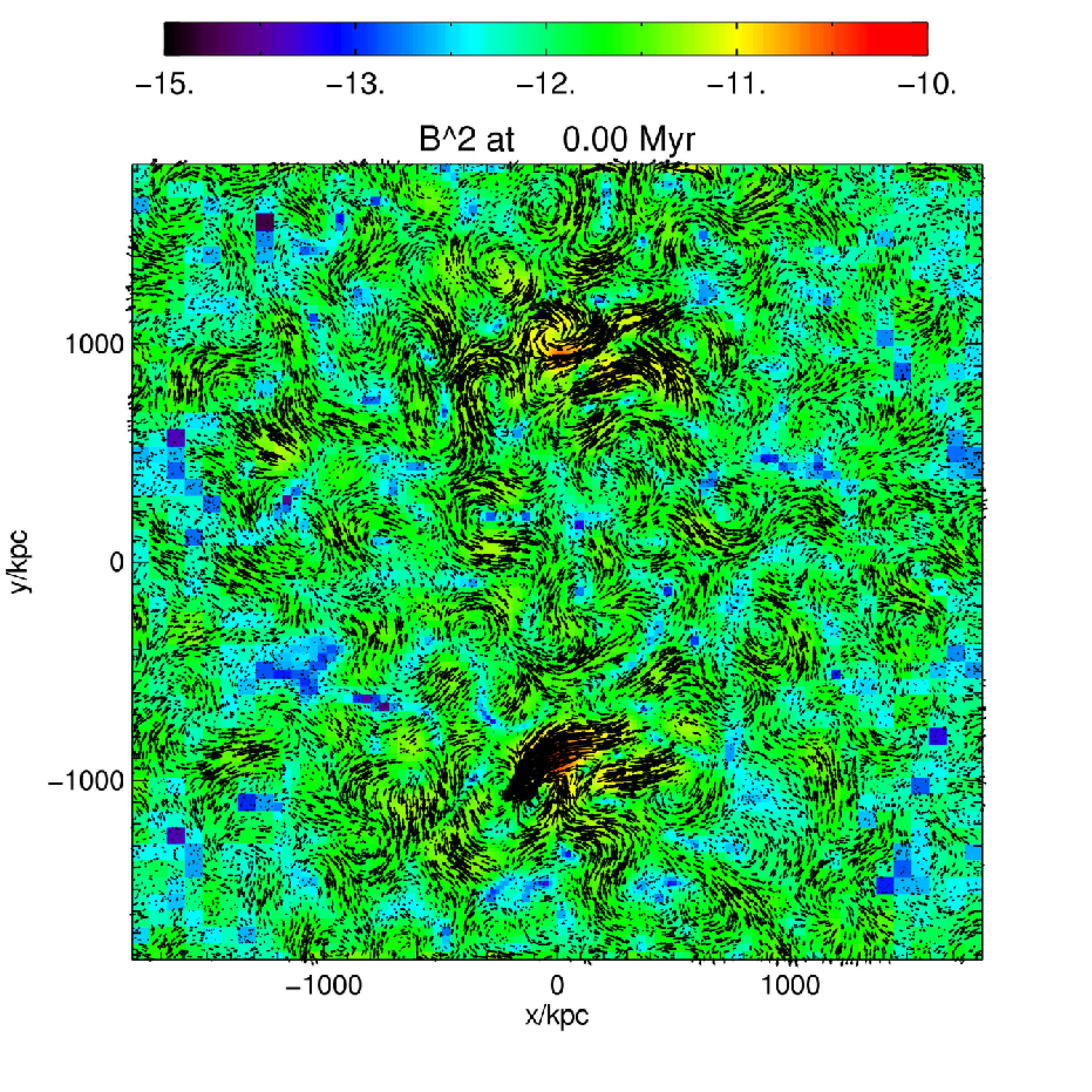}}
\centering{\includegraphics[width={0.49\columnwidth}]{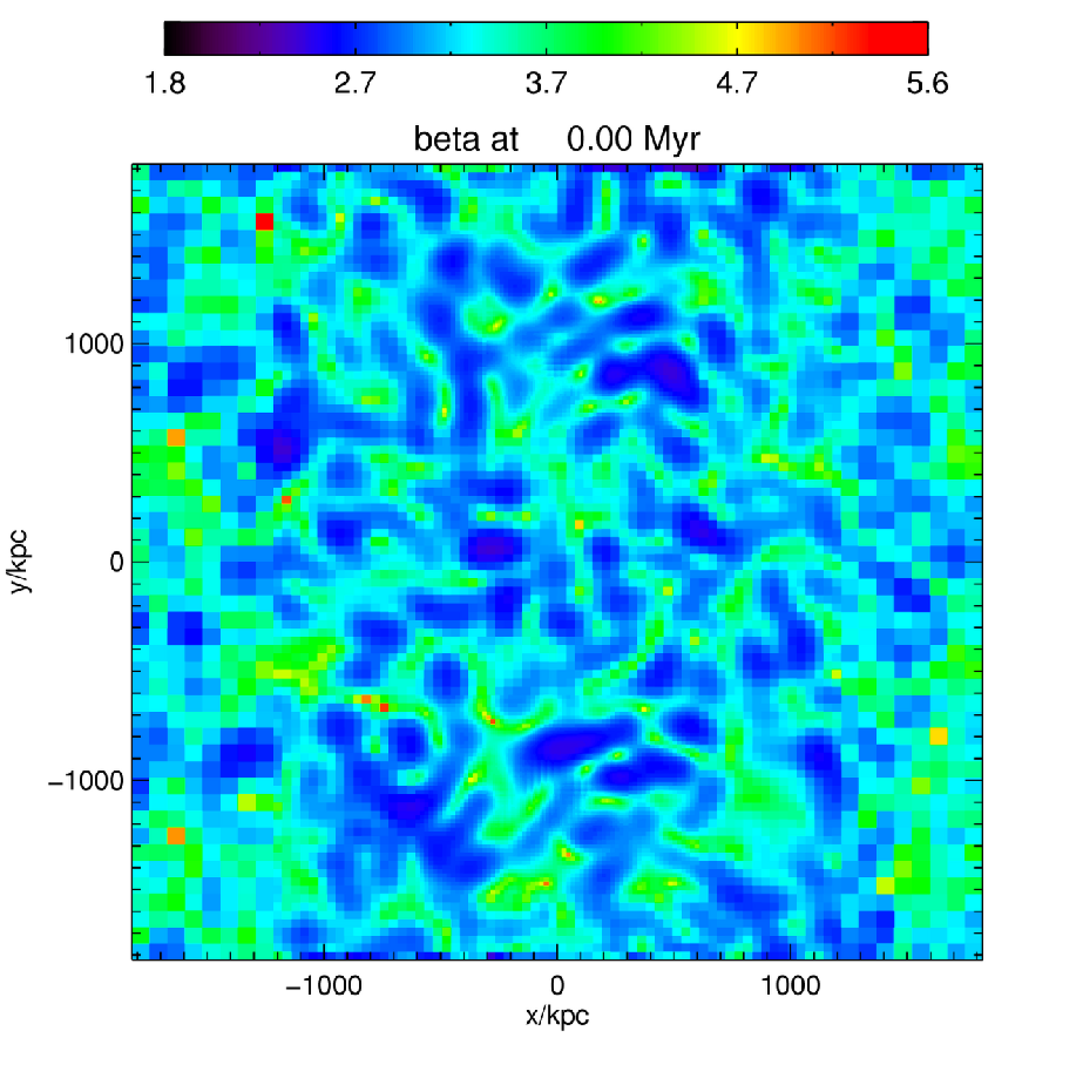}}
\centering{\includegraphics[width={0.49\columnwidth}]{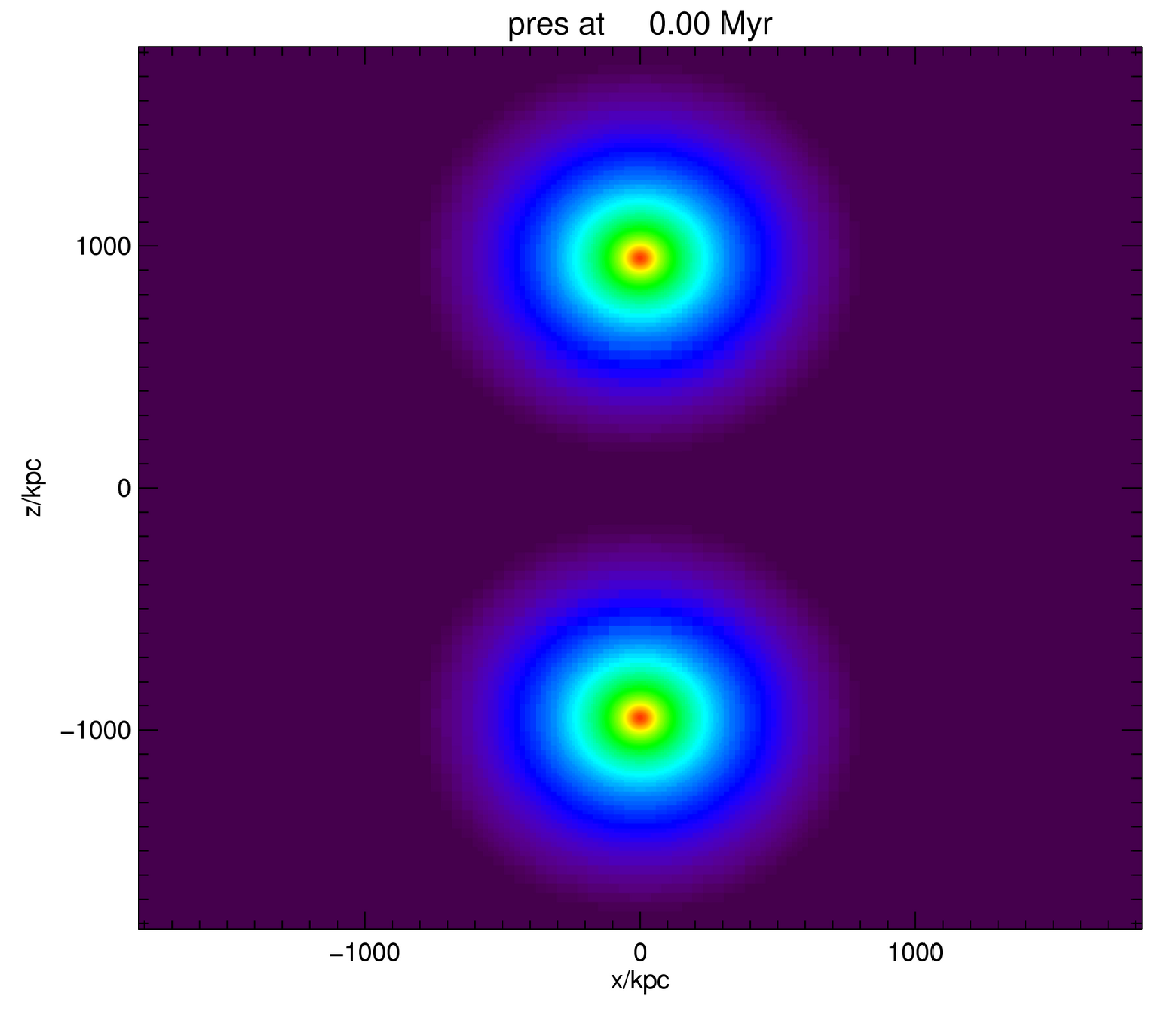}}
\caption{Initial conditions: slices of logarithmic gas densities, magnetic fields, plasma $\beta$ and pressure.}
\label{fig:init}
\end{figure}

\begin{figure}
\centering{\includegraphics[width={0.49\columnwidth}]{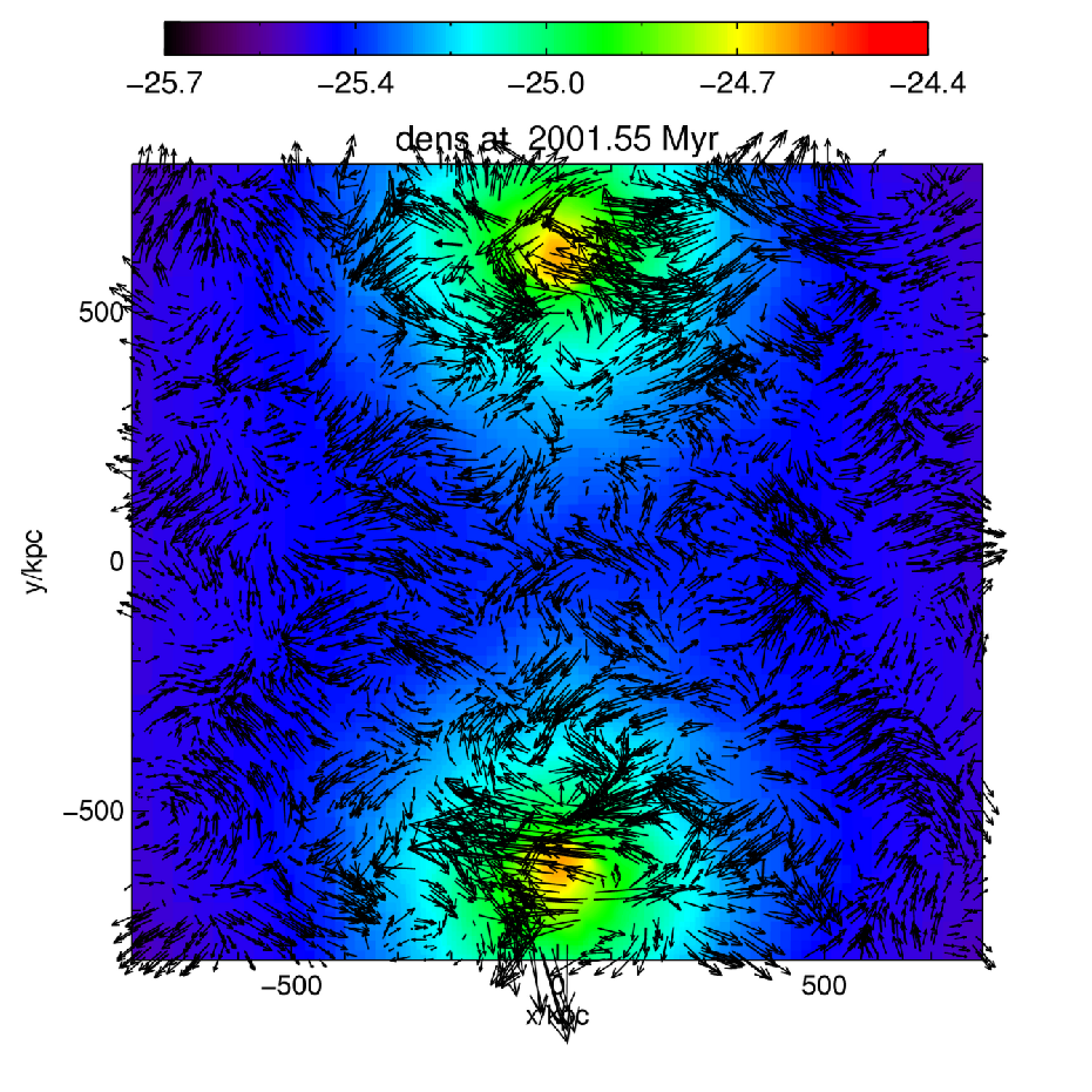}}
\centering{\includegraphics[width={0.49\columnwidth}]{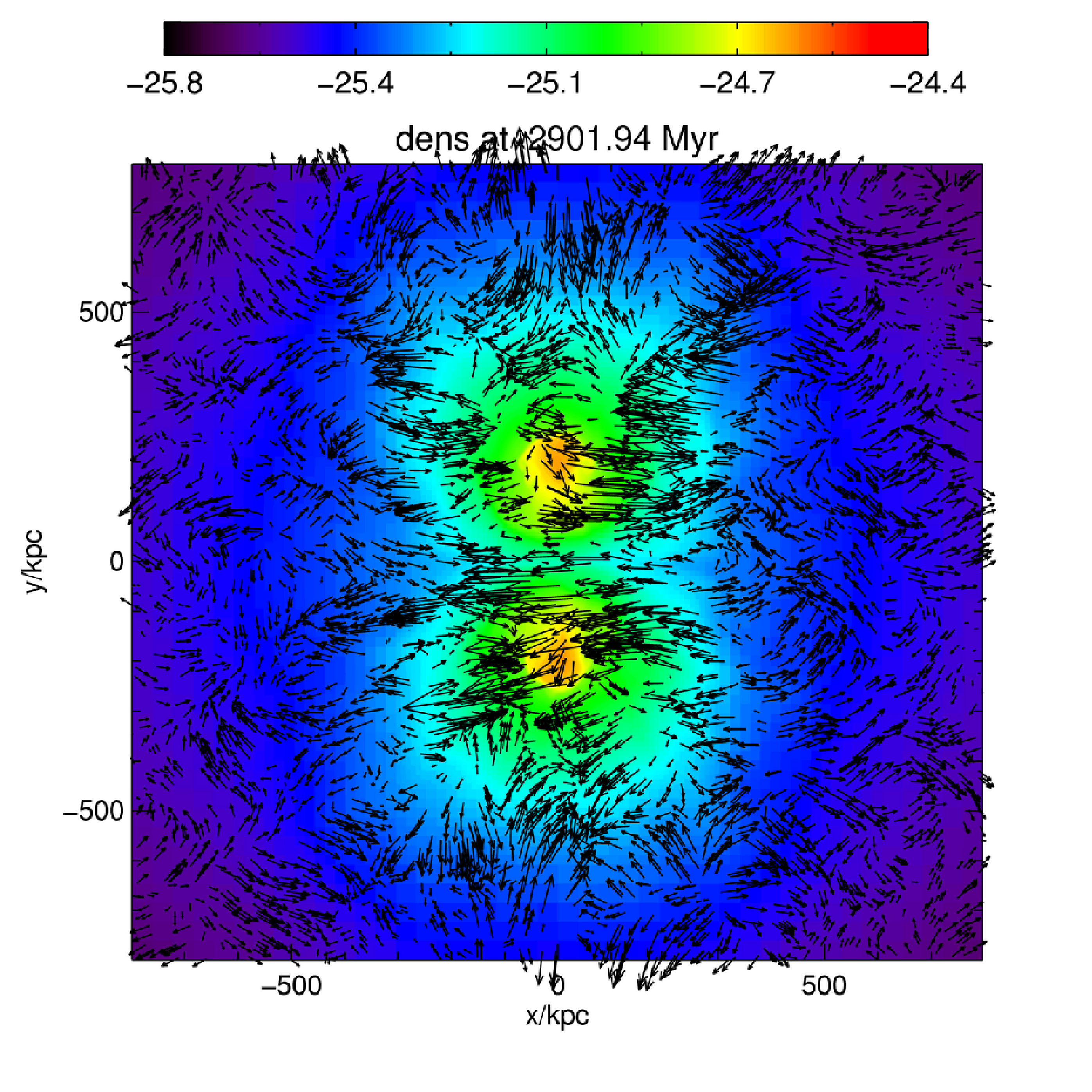}}
\centering{\includegraphics[width={0.49\columnwidth}]{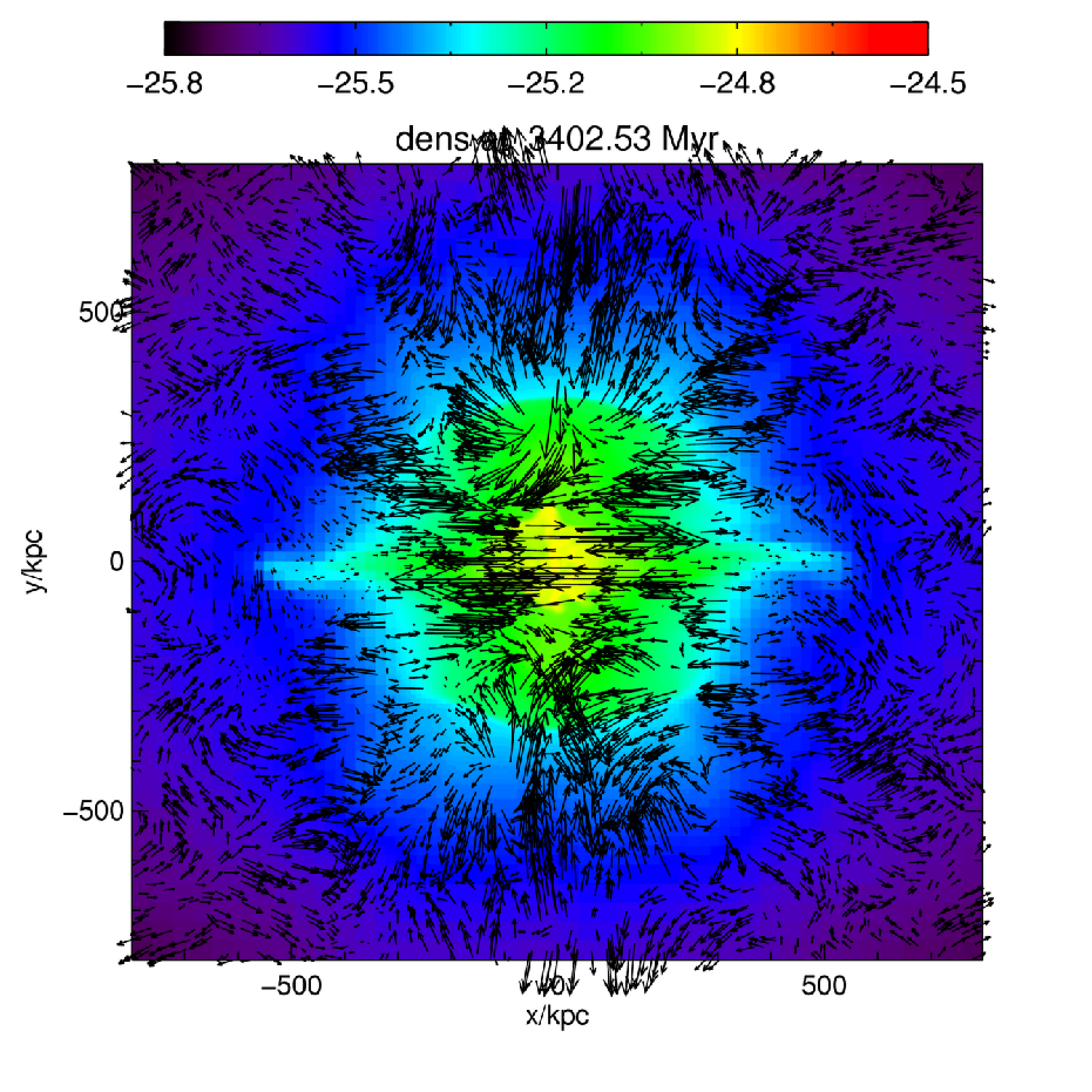}}
\centering{\includegraphics[width={0.49\columnwidth}]{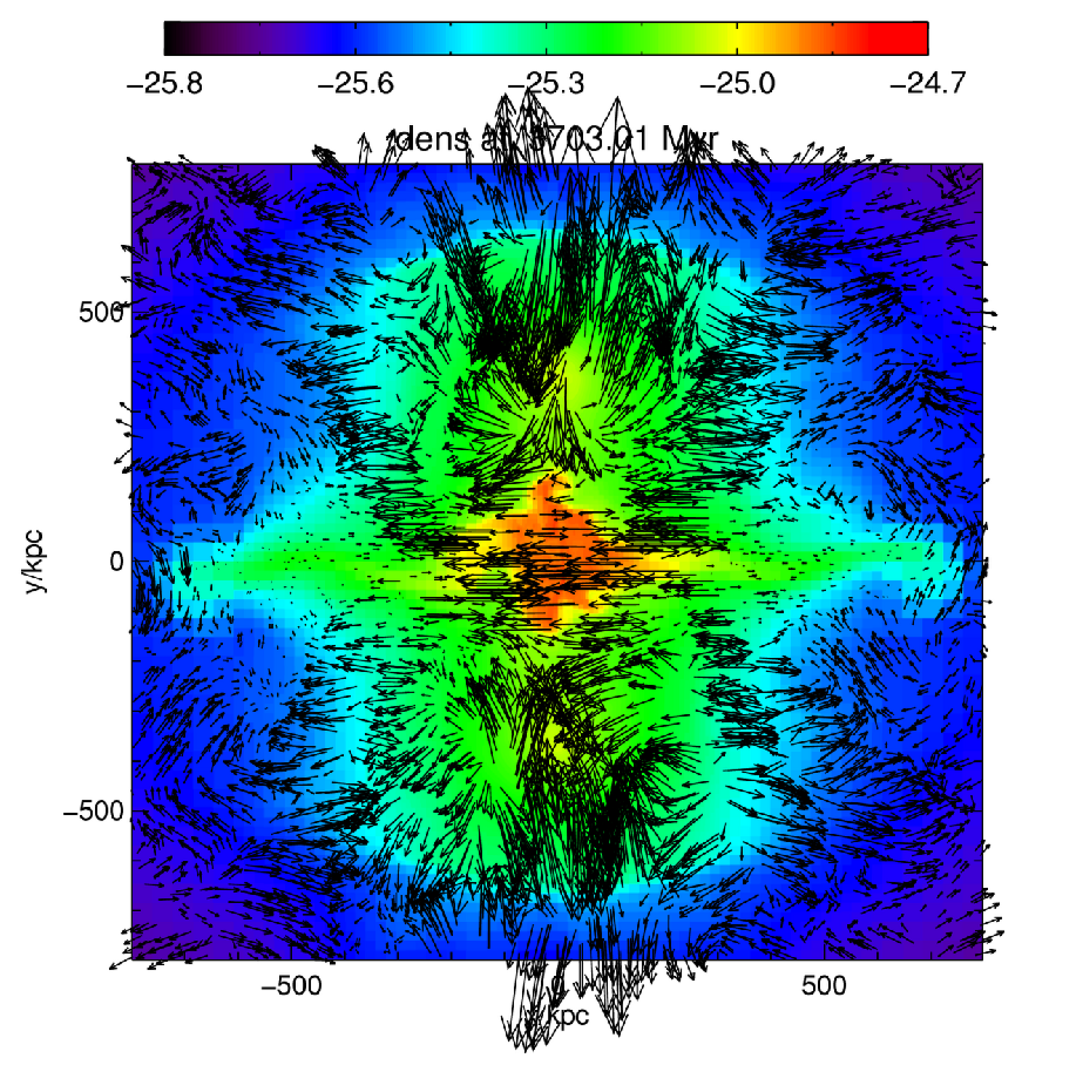}}
\caption{Slices of logarithmic gas densities (in a zoomed region) with magnetic field vectors superimposed for an equal-mass head-on collision.}
\label{fig:BdensR1}
\end{figure} 

\begin{itemize}

\item Initially the magnetic fields become filamentary in wake of galaxy clusters as result of stretching.
\item As the clusters approach each other: enhancement of density, temperature and magnetic field in the midplane magnetic field is perpendicular to the shock axis
\item After centre-passage: shock waves run diametrically out of the cluster. Magnetic field is perpendicular to shock front. 
\item Cluster gas dynamics gets dominated by large-scale flows and large eddies that shape the morphology of the cluster-wide field. Along the merger axis, the flow is largely directed outwards (away from the gravitational centres) along the axis. This results in magnetic fields that are perpendicular to the outgoing merger shock waves.
\item As the magnetic fields observed in the giant radio relic in CIZA J2242.8+530 lie within the shock plane, these magnetic fields are difficult to explain by a mere amplification of ambient magnetic fields.

\end{itemize}

These findings are consistent with what has been found in MHD simulations by \citep{ruszkowski10}.
They have performed full cosmological simulations of galaxy cluster formation that simultaneously include magnetic fields, radiative cooling 
and anisotropic thermal conduction. The motivation behind this work was that in isolated and idealized cluster models, the magnetothermal instability (MTI) tends to reorient the magnetic fields radially whenever the temperature gradient points in the direction opposite to gravitational acceleration.
Using cosmological simulations of the {\it Santa Barbara} cluster Ruszkowski. et al. (2010) have detected a radial bias in the velocity and magnetic fields. 
However, it was found that such radial bias is caused by the accretion of gas and substructure and not by MTI-driven field rearrangements.
The anisotropy effect is potentially detectable via radio polarization measurements with {\it LOFAR} and {\it Square Kilometer Array} 
and future X-ray spectroscopic studies with the {\it International X-ray Observatory}. 
Furthermore it was demonstrated that radiative cooling boosts the amplification of the magnetic field by about two orders of magnitude
beyond what is expected in the non-radiative cases. This effect is caused by the compression of the gas and frozen-in magnetic field 
as it accumulates in the cluster center. At $z=0$ the field is amplified by a factor of about $10^{6}$ compared to 
the uniform magnetic field evolved due to the universal expansion alone. 
Interestingly, the runs that include both radiative cooling and anisotropic thermal conduction exhibit stronger magnetic 
field amplification than purely radiative runs, especially at the off-center locations.
In these runs, shallow temperature gradients away from the cluster center 
make the ICM neutrally buoyant. Thus, the ICM is more easily mixed in these regions and the winding up 
of the frozen-in magnetic field is more efficient resulting in stronger magnetic field amplification.

\begin{acknowledgements}
MB acknowledges support by the research group FOR 1254 funded by the Deutsche Forschungsgemeinschaft. The software used in this work was in part developed by the DOE-supported ASC/Alliance Center for
Astrophysical Thermonuclear Flashes at the University of Chicago. RJvW acknowledges funding from the Royal  
Netherlands Academy of Arts and Sciences.
\end{acknowledgements}

\bibliographystyle{aa}

\bibliography{clusters}

\begin{thebibliography}{13}
\expandafter\ifx\csname natexlab\endcsname\relax\def\natexlab#1{#1}\fi

\bibitem[{{Burns} {et~al.}(1994){Burns}, {Roettiger}, {Ledlow}, \&
  {Klypin}}]{burns94}
{Burns}, J.~O., {Roettiger}, K., {Ledlow}, M., \& {Klypin}, A. 1994, \apjl,
  427, L87

\bibitem[{Dubey {et~al.}(2009)Dubey, Antypas, Ganapathy, Reid, Riley, Sheeler,
  Siegel, \& Weide}]{dubey09}
Dubey, A., Antypas, K., Ganapathy, M.~K., {et~al.} 2009, Parallel Computing,
  35, 512

\bibitem[{{Ensslin} {et~al.}(1998){Ensslin}, {Biermann}, {Klein}, \&
  {Kohle}}]{1998A&A...332..395E}
{Ensslin}, T.~A., {Biermann}, P.~L., {Klein}, U., \& {Kohle}, S. 1998, \aap,
  332, 395

\bibitem[{{Kocevski} {et~al.}(2007){Kocevski}, {Ebeling}, {Mullis}, \&
  {Tully}}]{2007ApJ...662..224K}
{Kocevski}, D.~D., {Ebeling}, H., {Mullis}, C.~R., \& {Tully}, R.~B. 2007,
  \apj, 662, 224

\bibitem[{{Miniati} {et~al.}(2001){Miniati}, {Jones}, {Kang}, \&
  {Ryu}}]{2001ApJ...562..233M}
{Miniati}, F., {Jones}, T.~W., {Kang}, H., \& {Ryu}, D. 2001, \apj, 562, 233

\bibitem[{{Pearce} {et~al.}(1994){Pearce}, {Thomas}, \& {Couchman}}]{pearce94}
{Pearce}, F.~R., {Thomas}, P.~A., \& {Couchman}, H.~M.~P. 1994, \mnras, 268,
  953

\bibitem[{{Ricker}(1998)}]{ricker98}
{Ricker}, P.~M. 1998, \apj, 496, 670

\bibitem[{{Ricker} \& {Sarazin}(2001)}]{ricker01}
{Ricker}, P.~M. \& {Sarazin}, C.~L. 2001, \apj, 561, 621

\bibitem[{{Roettiger} {et~al.}(1993){Roettiger}, {Burns}, \&
  {Loken}}]{roettiger93}
{Roettiger}, K., {Burns}, J., \& {Loken}, C. 1993, \apjl, 407, L53

\bibitem[{{Roettiger} {et~al.}(1997){Roettiger}, {Loken}, \&
  {Burns}}]{roettiger97}
{Roettiger}, K., {Loken}, C., \& {Burns}, J.~O. 1997, \apjs, 109, 307

\bibitem[{{Ruszkowski} {et~al.}(2010){Ruszkowski}, {Lee}, {Bruggen}, {Parrish},
  \& {Oh}}]{ruszkowski10}
{Ruszkowski}, M., {Lee}, D., {Bruggen}, M., {Parrish}, I., \& {Oh}, S.~P. 2010,
  ArXiv e-prints

\bibitem[{{Schindler} \& {Mueller}(1993)}]{schindler93}
{Schindler}, S. \& {Mueller}, E. 1993, \aap, 272, 137

\bibitem[{{van Weeren} {et~al.}(2010){van Weeren}, {R{\"o}ttgering},
  {Br{\"u}ggen}, \& {Hoeft}}]{2010Sci...330..347V}
{van Weeren}, R.~J., {R{\"o}ttgering}, H.~J.~A., {Br{\"u}ggen}, M., \& {Hoeft},
  M. 2010, Science, 330, 347

\end{thebibliography}


\end{document}